\documentclass[a4paper]{article}
\usepackage[11pt]{extsizes}
\usepackage[pdftex,colorlinks=true,linkcolor=blue,citecolor=blue,urlcolor=blue]{hyperref}
\usepackage{graphicx}
\usepackage{amsmath}
\usepackage{lscape}
\usepackage{geometry}

\begin{document}

\title{\textbf{A} \textbf{new large scale instability in rotating stratified fluids
driven by small scale forces} }
\author{Anatoly TUR, Malik CHABANE \\
Universit\'{e} de Toulouse [UPS],\\
CNRS, Institut de recherche en astrophysique et plan\'{e}tologie,\\
9 avenue du Colonel Roche, BP 44346, \\
31028 Toulouse Cedex 4, France \and Vladimir YANOVSKY \\
Institute for Single Crystals, National Academy of\\
Science Ukraine, Kharkov 31001, Ukraine}
\maketitle

\section*{Abstract}

\addcontentsline{toc}{section}{Abstract}

In this paper, we find a new large scale instability displayed by a
stratified rotating flow in forced turbulence. The turbulence is generated
by a small scale external force at low Reynolds number. The theory is built
on the rigorous asymptotic method of multi-scale development. There is no
other special constraint concerning the force. In previous papers, the force
was either helical or violating parity invariance. The nonlinear equations
for the instability are obtained at the third order of the perturbation
theory. In this article, we explain a detailed study of the linear stage of
the instability.

\bigskip

\bigskip Keywords: Large scale vortex instability, Coriolis forse, buoyancy,
multi-scale development, small scale turbulence.

Pacs: 47.32C-;47.27.De;47.27.em;47.55.Hd.

\section{Introduction}

\addcontentsline{toc}{section}{Introduction} Large scale instabilities are
very important in fluid dynamics. They generate vortices which play a
fundamental role in turbulence and in transport processes. The
characteristic dimensions of the large scale structures are greater than the
typical scale of the turbulence. The turbulence is often simulated using a
small scale external force. In this case, the large scale vortices are much
greater than the scale of the external force. Large scale vortices are well
observed in planetary atmospheres \cite{[11]}, \cite{[12]}, in numerical
simulations, and in laboratory experiments \cite{[2]}, \cite{[8]}. The
generation process of large scale instabilities has been studied in several
papers \cite{[15]}, \cite{[20]}. In these papers, the turbulence which
generates these coherent large scale structures can not be homogenous,
isotropic, or mirror invariant. A series of papers have shown that the
essential mechanism which leads to the generation of large scale vortices is
the lack of reflection invariance. This mechanism was called the
hydrodynamic $\alpha $-effect by analogy with the similar mechanism of
generation of large scale magnetic fields.\newline
Turbulence lacking reflection invariance is helical and a pseudo-scalar $%
\vec{v}\cdot \vec{rot}\vec{v}\neq 0$ appears. Nevertheless, the helicity of
turbulence by itself can not generate large scale vortices. Other factors
which lack reflection invariance are necessary, such as, for instance,
compressibility \cite{[17]}, \cite{[20]} or temperature gradients \cite{[18]}%
, \cite{[19]}. Large scale instability can also appear if the turbulence
lacks parity invariance (AKA effect)\cite{[16]}. The helicity of the
turbulence can be defined in a phenomenological way, but helicity can also
be generated by an internal mechanism like rotation or buoyancy \cite{[16a]}%
, \cite{[16c]}, \cite{[27]}. \newline
Large scale instability in a stratified rotating flow was studied in \cite
{[24]}, \cite{[25]}. In \cite{[24]} it was shown that a rotating
incompressible flow with a constant temperature gradient can not display a
large scale instability. In \cite{[25]} the author presented large scale
instabilities with a quadratic temperature gradient. In both papers, the
authors used the functional averaging method. This method has some
inconveniences. Especially it is impossible to make a strict hierarchy of
orders as in perturbation theory. This means that it is impossible to
identify the orders in which the instability appears and the ones in which
it is absent. That is why the fact that the instability is absent when using
the functional averaging method can not exclude its occurrence when using
the rigorous asymptotic method of multi-scale development.\newline
The occurrence of large scale instability in helical stratified turbulence
was confirmed by the multi-scale development method in \cite{[26]}. In that
paper it was shown that the instability appears at the third order in the
asymptotical development built on the small value of the Reynolds number.
But in the first papers on this subject, using the functional averaging
method, it was not clear in which order the instability would appear.\newline
Direct numerical simulation of the Boussinesq equation confirmed the
existence of large scale vortex generation in stratified and rotating flows 
\cite{[14a]}, \cite{[14b]}. Sometimes the appearance of large scale vortex
structures is accompanied by an inverse cascade of energy both in the
three-dimensional case (AKA-effect \cite{[12a]}), and in the quasi
two-dimensional case \cite{[3]}, \cite{[3c]}, \cite{[7]}, \cite{[8]}. One
may say that the inverse cascade itself is also one of the mechanism of the
generation of large scale structures \cite{[3a]}, \cite{[23]}. One of the
important large scale instabilities in an incompressible fluid is the AKA
effect (Anisotropic Kinetic Alpha effect) which was found in the work of
Frisch, She and Sulem \cite{[15]}. In this paper, the large scale
instability appears under the impact of a small scale force in which parity
is broken (with zero helicity). In a later paper \cite{[16]}, the inverse
cascade of energy and the nonlinear mode of instability saturation were
studied. Despite the fact that the broken parity is a more general notion
than helicity, in fact, the helicity $\vec{v}\cdot rot\vec{v}\neq 0$ is the
widespread mechanism of symmetry breaking in hydrodynamical flows. The
injection of a helical external force into a hydrodynamic system has been
studied in several papers (\cite{[17]}, \cite{[20]}). As a result, it was
understood that a small scale turbulence that is able to generate large
scale perturbations can not be simply homogeneous, isotropic, and helical 
\cite{[14]}, but must have additional special properties. In some cases, the
existence of a large scale instability has been shown (a vortex dynamo or
the hydrodynamic $\alpha $-effect). In the magneto hydrodynamics of a
conductive fluid, the $\alpha $-effect is well known \cite{[13]}. In
particular, in \cite{[18]} it was shown that a large scale instability
exists in convective systems with small scale helical turbulence. These
papers as well as the results of numerical modelling are described in detail
in the review article \cite{[21]}, which is focused essentially on possible
applications of these results to the issue of tropical cyclone origination.
In this paper, we develop an analytical theory of the new large scale
instability which generates large scale vortices in a stratified rotating
flow with a constant temperature gradient under the action of a small scale
external force which does not have any particular properties (especially it
is nonhelical and it does not lack parity invariance). The force only
maintains turbulent fluctuations. In other words, this force can not display
any instability. But the situation changes when both the Coriolis force and
the buoyancy are added to this force. The joint action of these forces
generates an internal helicity, which in turn generates an instability. The
theory of this instability is developed rigourously using the method of
asymptotic multi-scale development similar to what was done by Frisch, She
and Sulem for the theory of the AKA effect \cite{[15]}. This method allows
finding the equations for large scale perturbations as the secular equations
of perturbation theory, to calculate the Reynolds stress tensor and to find
the instability. Our paper is organised as follows: in Section 2 we
formulate the problem and the equations for the Coriolis force and the
stratification in the Boussinesq approximation; in Section 3 we examine the
principal scheme of the multi-scale development and we give the secular
equations. In Section 4 we describe the calculations of the Reynolds stress.
In Section 5 we discuss the instability and the conditions for its
realization. The results obtained are discussed in the conclusions given in
Section 6. The Reynolds stress and internal helicity are calculted in
Appendices A and B, respectively.

\bigskip

\section{The main equations and formulation of the problem}

Let us consider the equations for the motion of an incompressible fluid with
a constant temperature gradient in the Boussinesq approximation:

\begin{equation}
\frac{\partial \vec{V}}{\partial t}+(\vec{V}\cdot \nabla )\vec{V}+2\vec{%
\Omega}\times \vec{V}=-\frac{1}{\rho _{0}}\nabla P+\nu \Delta \vec{V}+g\beta
T\vec{l}+\vec{F}_{0}\newline
,  \label{1}
\end{equation}

\begin{equation}
\frac{\partial T}{\partial t}+(\vec{V}\cdot \nabla )T=\chi \Delta T-V_{z}A,
\label{2}
\end{equation}

\begin{equation}
\nabla \cdot \vec{V}=0.
\end{equation}

Here, $\vec{l}=\left( 0,0,1\right) $, $\beta $ is the thermal expansion
coefficient, $A=\frac{dT_{0}}{dz}$ is the constant equilibrium gradient of
the temperature, $\rho _{0}=Const.$, and $\nabla T_{0}=A\vec{l}$. The
external force $\vec{F}_{0}$ has zero divergence. Let $\lambda
_{0},t_{0},f_{0},v_{0}$ be, respectively, the characteristic scale, time,
amplitude of the external force, and velocity of our system. We choose the
dimensionless variables $(t,\vec{x},\vec{V})$:

\begin{equation*}
\vec{x}\rightarrow \frac{\vec{x}}{\lambda _{0}},t\rightarrow \frac{t}{t_{0}},%
\vec{V}\rightarrow \frac{\vec{V}}{v_{0}},\vec{F}_{0}\rightarrow \frac{\vec{F}%
_{0}}{f_{0}},P\rightarrow \frac{P}{\rho _{0}P_{0}},
\end{equation*}

\begin{equation*}
t_{0}=\frac{\lambda _{0}^{2}}{\nu },P_{0}=\frac{v_{0}\nu }{\lambda _{0}}%
,f_{0}=\frac{v_{0}\nu }{\lambda _{0}^{2}},v_{0}=\frac{f_{0}\lambda _{0}^{2}}{%
\nu }.
\end{equation*}

Then,

\begin{equation*}
\frac{\partial \vec{V}}{\partial t}+R(\vec{V}\cdot \nabla )\vec{V}+D\vec{l}%
\times \vec{V}=-\nabla P+\Delta \vec{V}+(\frac{\lambda _{0}^{2}}{v_{0}\nu }%
)g\beta T\vec{l}+\vec{F}_{0},
\end{equation*}

\begin{equation*}
\frac{\partial T}{\partial t}+R(\vec{V}\cdot \nabla )T=\frac{1}{Pr}\Delta
T-RV_{z}(A\lambda _{0}),
\end{equation*}

\noindent where $R=\frac{\lambda _{0}v_{0}}{\nu },D=\sqrt{Ta}$ where $R$ and 
$Ta=\frac{4{\Omega }^{2}\lambda _{0}^{4}}{{\nu }^{2}}$ are respectively the
Reynolds number and the Taylor number on scale $\lambda _{0}$. $Pr=\frac{\nu 
}{\chi }$ represents the Prandtl number. We introduce the dimensionless
temperature $T\rightarrow \frac{T}{\lambda _{0}A}$, and obtain the system of
equations

\begin{equation*}
\frac{\partial \vec{V}}{\partial t}+R(\vec{V}\cdot\nabla ) \vec{V}-\Delta 
\vec{V}+D\vec{l}\times\vec{V}=-\nabla P+\frac{Ra}{RPr }T \vec{l}+\vec{F}_{0}
\end{equation*}

\begin{equation*}
\frac{1}{R}\left( \frac{\partial T}{\partial t}-\frac{1}{Pr }\Delta T\right)
=-V_{z}-(\vec{V}\cdot\nabla )T.
\end{equation*}

Here, $Ra=\frac{\lambda _{0}^{4}Ag\beta }{\chi \nu }$ is the Rayleigh number
on the scale $\lambda _{0}$. Furthermore, for the purpose of simplification,
we will consider the case $Pr =1$. We pass to the new temperature $%
T\rightarrow \frac{T}{R}$, and obtain

\begin{equation}
\frac{\partial \vec{V}}{\partial t}+R(\vec{V}\cdot \nabla )\vec{V}-\Delta 
\vec{V}+D\vec{l}\times \vec{V}=-\nabla P+RaT\vec{l}+\vec{F}_{0},  \label{3}
\end{equation}

\begin{equation}
\left( \frac{\partial T}{\partial t}-\Delta T\right) =-V_{z}-R(\vec{V}\cdot
\nabla )T,  \label{4}
\end{equation}

\begin{equation*}
\nabla\cdot\vec{V}=0.
\end{equation*}

We will consider as a small parameter of an asymptotic development the
Reynolds number $R=\frac{\lambda _{0}v_{0}}{\nu }$ $\ll 1$ on the scale $%
\lambda _{0}$. Concerning the parameters $Ra$ and $D$, we do not choose any
range of values for the moment. Let us examine the following formulation of
the problem. We consider the external force as being small and of high
frequency. This force leads to small scale fluctuations in velocity and
temperature against a background of equilibrium. After averaging, these
quickly oscillating fluctuations vanish. Nevertheless, due to small
nonlinear interactions in some orders of perturbation theory, nonzero terms
can occur after averaging. This means that they are not oscillatory, that is
to say, they are large scale. From a formal point of view, these terms are
secular, i.e., they create the conditions for the solvability of a large
scale asymptotic development. So the purpose of this paper is to find and
study the solvability equations, i.e., the equations for large scale
perturbations. Let us denote the small scale variables by $x_{0}=(\vec{x}%
_{0},t_{0})$ , and the large scale ones by $X=(\vec{X},T)$. The small scale
partial derivative operation $\frac{\partial }{\partial x_{0}^{i}},\frac{%
\partial }{ \partial t_{0}}$, and the large scale ones $\frac{\partial }{%
\partial \vec{X}}, \frac{\partial }{\partial T}$ are written, respectively,
as $\partial _{i},\partial _{t},\nabla _{i}$ and $\partial _{T}$. To
construct a multi-scale asymptotic development we follow the method which is
proposed in \cite{[15]}.

\section{The multi-scale asymptotic development}

Let us search for the solution to Equations (\ref{3}) and (\ref{4}) in the
following form: 
\begin{equation}
\vec{V}(\vec{x},t)=\frac{1}{R}\vec{W}_{-1}(X)+\vec{v}_{0}(x_{0})+R\vec{v}%
_{1}+R^{2}\vec{v}_{2}+R^{3}\vec{v}_{3}+\cdots ,  \label{68.1}
\end{equation}

\begin{equation}
T(\vec{x},t)=\frac{1}{R}T_{-1}(X)+T_{0}(x_{0})+RT_{1}+R^{2}T_{2}+R^{3}T_{3}+%
\cdots ,  \label{68.2}
\end{equation}

\begin{equation}
P(\vec{x},t)=\frac{1}{R^{3}}P_{-3}(X)+\frac{1}{R^{2}}P_{-2}(X)+\frac{1}{R}%
P_{-1}(X)+P_{0}(x_{0})+R(P_{1}+\overline{P}_{1}(X))+R^{2}P_{2}+R^{3}P_{3}+%
\cdots .
\end{equation}

Let us introduce the following equalities: $\vec{X}=R^2\vec{x}_{0}$ and $%
T=R^4t_{0}$ which lead to the expression for the space and time derivatives:

\begin{equation}
\frac{\partial }{\partial x^{i}}=\partial _{i}+R^{2}\nabla _{i},  \label{5}
\end{equation}

\begin{equation}
\frac{\partial }{\partial t}=\partial _{t}+R^{4}\partial _{T},  \label{6}
\end{equation}

\begin{equation}
\frac{\partial ^{2}}{\partial {x^{j}}\partial {x^{j}}}=\partial
_{jj}+2R^{2}\partial _{j}\nabla _{j}+R^{4}\partial _{jj}.  \label{7}
\end{equation}

Using indicial notation, the system of equation can be written as

\begin{equation}
(\partial _{t}+R^{4}\partial _{T})V^{i}+R(\partial _{j}+R^{2}\nabla
_{j})(V^{i}V^{j})+D\varepsilon _{ijk}l^{j}V^{k}=(\partial _{i}+R^{2}\nabla
_{i})P+(\partial _{jj}+2R^{2}\partial _{j}\nabla _{j}+R^{4}\nabla
_{jj})V^{i}+RaTl^{i}+F_{0}^{i},
\end{equation}

\begin{equation}
\partial _{t}T-\partial _{jj}T=-V^{z}-R\partial _{j}\left( V^{i}T\right) ,
\end{equation}

\begin{equation}
\left( \partial _{i}+R^{2}\nabla _{i}\right) V^{i}=0.
\end{equation}

Substituting these expressions into the initial equations (\ref{3}) and (\ref
{4}) and then gathering together the terms of the same order, we obtain the
equations of the multi-scale asymptotic development and write down the
obtained equations up to order $R^{3}$ inclusive. In the order $R^{-3}$
there is only the equation

\begin{equation}
\partial _{i}P_{-3}=0,\Rightarrow P_{-3}=P_{-3}(X).  \label{D1}
\end{equation}

In order $R^{-2}$ we have the equation

\begin{equation}
\partial _{i}P_{-2}=0,\Rightarrow P_{-2}=P_{-2}(X).  \label{D2}
\end{equation}

In order $R^{-1}$ we get a system of equations:

\begin{equation}
\partial _{t}W_{-1}^{i}-\partial _{jj}W_{-1}^{i}+D\varepsilon
_{ijk}l^{j}W_{-1}^{k}=-(\partial _{i}P_{-1}+\nabla
_{i}P_{-3})+RaT_{-1}l^{i}-\partial _{j}W_{-1}^{i}W_{-1}^{j},  \label{D3}
\end{equation}

\begin{equation}
\partial _{t}T_{-1}-\partial _{jj}T_{-1}=-\partial
_{j}W_{-1}^{j}T_{-1}-W_{-1}^{z},  \label{D4}
\end{equation}

\begin{equation*}
\partial _{i}W_{-1}^{i}=0.
\end{equation*}

The system of equations (\ref{D3}) and (\ref{D4}) gives the secular terms

\begin{equation}
-\nabla _{i}P_{-3}+RaT_{-1}l^{i}=D\varepsilon _{ijk}l^{j}W_{-1}^{k},
\label{D5}
\end{equation}
which corresponds to a geostrophic equilibrum equation, and 
\begin{equation}
W_{-1}^{z}=0.  \label{D6}
\end{equation}

In zero order $R^{0}$, we have the following system of equations:

\begin{equation}
\partial _{t}v_{0}^{i}-\partial _{jj}v_{0}^{i}+\partial
_{j}(W_{-1}^{i}v_{0}^{j}+v_{0}^{i}W_{-1}^{j})+D\varepsilon
_{ijk}l^{j}v_{0}^{k}=-(\partial _{i}P_{0}+\nabla
_{i}P_{-2})+RaT_{0}l^{i}+F_{0}^{i},  \label{D7}
\end{equation}

\begin{equation}
\partial _{t}T_{0}-\partial _{jj}T_{0}+\partial
_{j}(W_{-1}^{j}T_{0}+v_{0}^{j}T_{-1})=-v_{0}^{z},  \label{D8}
\end{equation}

\begin{equation*}
\partial_{i} v_{0}^{i}=0.
\end{equation*}

These equations give one secular equation:

\begin{equation}
\nabla P_{-2}=0,\Rightarrow P_{-2}=Const.  \label{D9}
\end{equation}

Let us consider the equations of the first approximation $R$:

\begin{equation}
\partial _{t}v_{1}^{i}-\partial _{jj}v_{1}^{i}+D\varepsilon
_{ijk}l^{j}v_{1}^{k}+\partial
_{j}(W_{-1}^{i}v_{1}^{j}+v_{1}^{i}W_{-1}^{j}+v_{0}^{i}v_{0}^{j})=-\nabla
_{j}(W_{-1}^{i}W_{-1}^{j})-(\partial _{i}P_{1}+\nabla
_{i}P_{-1})+RaT_{1}l^{i},  \label{D10}
\end{equation}

\begin{equation}
\partial _{t}T_{1}-\partial _{jj}T_{1}+\partial
_{j}(W_{-1}^{j}T_{1}+v_{1}^{j}T_{-1}+v_{0}^{j}T_{0})+\nabla
_{j}(W_{-1}^{j}T_{-1})=-v_{1}^{z},  \label{D11}
\end{equation}

\begin{equation}
\partial_{i} V_{1}^{i}+\nabla_{i} W_{-1}^{i}=0.  \label{D12}
\end{equation}

>From this system of equations there follows the secular equations:

\begin{equation}
\nabla _{i}W_{-1}^{i}=0,  \label{D13}
\end{equation}

\begin{equation}
\nabla _{j}(W_{-1}^{i}W_{-1}^{j})=-\nabla _{i}P_{-1},  \label{D14}
\end{equation}

\begin{equation}
\nabla _{j}(W_{-1}^{j}T_{-1})=0.  \label{D15}
\end{equation}

The secular equations (\ref{D13}) and (\ref{D15}) are satisfied by choosing
the following geometry for the velocity field:

\begin{equation}
\vec{W}=(W_{-1}^{x}(Z),W_{-1}^{y}(Z),0);T_{-1}=T_{-1}(Z);  \label{D16}
\end{equation}

\begin{equation*}
\nabla P_{-1}=0,\Rightarrow P_{-1}=Const.
\end{equation*}

In the second order $R^{2}$, we obtain the equations

\begin{equation}
\partial _{t}v_{2}^{i}-\partial _{jj}v_{2}^{i}-2\partial_{j} \nabla_{j}
v_{0}^{i}+\partial_{j}
(W_{-1}^{i}v_{2}^{j}+v_{2}^{i}W_{-1}^{j}+v_{0}^{i}v_{1}^{j}+v_{1}^{i}v_{0}^{j})+ D\varepsilon_{ijk}l^{j}v_{2}^{k}=
\label{D17}
\end{equation}

\begin{equation*}
=-\nabla _{j}(W_{-1}^{i}v_{0}^{j}+v_{0}^{i}W_{-1}^{j})-(\partial
_{i}P_{2}+\nabla _{i}P_{0})+RaT_{2}l^{i},
\end{equation*}

\begin{equation}
\partial _{t}T_{2}-\partial _{jj}T_{2}-2\partial _{j}\nabla
_{j}T_{0}+\partial
_{j}(W_{-1}^{j}T_{2}+v_{2}^{j}T_{-1}+v_{0}^{j}T_{1}+v_{1}^{j}T_{0})=-\nabla
_{j}(W_{-1}^{j}T_{0}+v_{0}^{j}T_{-1})-v_{2}^{z},  \label{D18}
\end{equation}

\begin{equation}
\partial_{i} v_{2}+\nabla_{i} v_{0}=0.  \label{D19}
\end{equation}

It is easy to see that there are no secular terms in this order..

Let us come now to the most important order $R^{3}$. In this order we obtain
the equations

\begin{equation}
\partial _{t}v_{3}^{i}+\partial _{T}W_{-1}^{i}-(\partial
_{jj}v_{3}^{i}+2\partial_{j} \nabla_{j} v_{1}^{i}+\nabla_{jj}
W_{-1}^{i})+\nabla_{j}
(W_{-1}^{i}v_{1}^{j}+v_{1}^{i}W_{-1}^{j}+v_{0}^{i}v_{0}^{j})+  \label{D20}
\end{equation}

\begin{equation*}
+\partial
_{j}(W_{-1}^{i}v_{3}^{j}+v_{3}^{i}W_{-1}^{j}+v_{0}^{i}v_{2}^{j}+v_{2}^{i}v_{0}^{j}+v_{1}^{i}v_{1}^{j})+D\varepsilon _{ijk}l^{j}v_{3}^{k}=-(\partial _{i}P_{3}+\nabla _{i}%
\overline{P}_{1})+RaT_{3}l^{i},
\end{equation*}

\begin{equation}
\partial _{t}T_{3}+\partial _{T}T_{-1}-(\partial _{jj}T_{3}+2\partial_{j}
\nabla_{j} T_{1}+\nabla_{jj} T_{-1})+\nabla_{j}
(W_{-1}^{j}T_{1}+v_{1}^{j}T_{-1}+v_{0}^{j}T_{0})+  \label{D21}
\end{equation}

\begin{equation*}
+\partial
_{j}(W_{-1}^{j}T_{3}+v_{3}^{j}T_{-1}+v_{0}^{j}T_{2}+v_{2}^{j}T_{0}+v_{1}^{j}T_{1})=-v_{3}^{z},
\end{equation*}

\begin{equation*}
\partial _{i}v_{3}+\nabla _{i}v_{1}=0.
\end{equation*}

>From this we get the main secular equation:

\begin{equation}
\partial _{T}W_{-1}^{i}-\Delta W_{-1}^{i}+\nabla _{k}(\overline{%
v_{0}^{k}v_{0}^{i}})=-\nabla _{i}\overline{P}_{1},  \label{68.4}
\end{equation}

\begin{equation}
\partial _{T}T_{-1}-\Delta T_{-1}+\nabla _{k}(\overline{v_{0}^{k}T_{0}})=0.
\label{68.5}
\end{equation}

There is also an equation to find the pressure $P_{-3}$:

\begin{equation}
-\nabla _{i}P_{-3}+RaT_{-1}l^{i}=D\varepsilon _{ijk}l^{j}W_{-1}^{k}.
\end{equation}

\section{Calculations of the Reynolds stresses}

It is clear that the essential equation for finding the nonlinear
alpha-effect is Equation (\ref{68.4}). In order to obtain these equations in
closed form, we need to calculate the Reynolds stresses $\nabla _{k}(%
\overline{v_{0}^{k}v_{0}^{i}})$. First of all we have to calculate the
fields of zero approximation $v_{0}^{k}$. From the asymptotic development in
zero order we have

\begin{equation}
\partial _{t}v_{0}^{i}-\partial _{jj}v_{0}^{i}+W_{-1}^{k}\partial
_{k}v_{0}^{i} + D\varepsilon_{ijk}l^{j}v_{0}^{k}=-\partial
_{i}P_{0}+RaT_{0}l^{i}+F_{0}^{i},  \label{68.11}
\end{equation}

\begin{equation}
\partial
_{t}T_{0}-\partial_{jj}T_{0}+W_{-1}^{k}\partial_{k}T_{0}=-v_{0}^{k}l^{k}.
\label{68.12}
\end{equation}

Let us introduce the operator $\widehat{D}_{0}$:

\begin{equation}
\widehat{D}_{0}\equiv \partial _{t}-\partial _{jj}+W^{k}\partial _{k}.
\label{68.13}
\end{equation}

Using $\widehat{D}_{0}$, we rewrite Equations (\ref{68.11}) and (\ref{68.12}%
):

\begin{equation}
\widehat{D}_{0}v_{0}^{i}+ D\varepsilon_{ijk}l^{j}v_{0}^{k}=-\partial
_{i}P_{0}+RaT_{0}l^{i}+F_{0}^{i},  \label{68.14}
\end{equation}

\begin{equation}
\widehat{D}_{0}T_{0}=-v_{0}^{k}l^{k}.  \label{68.15}
\end{equation}

Eliminating the temperature and pressure from Equation (\ref{68.14}), we
obtain

\begin{equation}
\left(\widehat{D}_{0}^{2}\delta _{ik}+\widehat{P}_{ip}Ral^{p}l^{k} + D%
\widehat{D}_{0}\widehat{P}_{ip}\varepsilon_{pjk}l^{j}\right)v_{0}^{k}= 
\widehat{D}_{0}F_{0}^{i}.  \label{68.19}
\end{equation}

Here, $\widehat{P}_{ip}$ is the projection operator

\begin{equation*}
\widehat{P}_{ip}\equiv \delta _{ip}-\frac{\partial _{i}\partial _{p}}{%
\partial _{jj}}.
\end{equation*}

Dividing this equation by $\widehat{D}_{0}^{2}$, we can write it in the form

\begin{equation}
M_{ik}v_{0}^{k}=\frac{F_{0}^{i}}{\widehat{D}_{0}}, ,  \label{68.20}
\end{equation}

\noindent where $M_{ik}$ is the operator given by

\begin{equation}
M_{ik}\equiv \delta _{ik}+Ra\frac{\widehat{P}_{ip}}{\widehat{D}_{0}^{2}}
l^{p}l^{k} + D\frac{\widehat{P}_{ip}}{\widehat{D}_{0}}\varepsilon_{pjk}l^{j}.
\end{equation}

We must now determine the inverse operator $M_{kj}^{-1}, i.e., $: $%
M_{ik}M_{kj}^{-1}=\delta _{ij}.$

After some calculation, we find

\begin{equation}
M_{kj}^{-1}= \frac{1}{\Psi}\left[\left(1+\frac{Ra}{\widehat{D}_{0}^2}%
\widehat{P}_{33}\right)\delta _{kj}+D\frac{\widehat{P}_{kp}}{\widehat{D}_{0}}%
\varepsilon_{pjn}l^{n}-Ra\frac{\widehat{P}_{kp}}{\widehat{D}_{0}^2}%
l^{p}l^{j}+\xi_{kj}\right].
\end{equation}

\noindent Here,

\begin{equation}
\Psi= 1 + \frac{D^2}{\widehat{D}_{0}^2}\widehat{P}_{11}+\frac{Ra}{\widehat{D}%
_{0}^2}\widehat{P}_{33}+\frac{D^2Ra}{\widehat{D}_{0}^4}\left(\widehat{P}_{11}%
\widehat{P}_{33}-\widehat{P}_{13}^2\right)=det(M)
\end{equation}

\noindent and

\begin{align*}
\xi= 
\begin{pmatrix}
0 & \frac{DRa}{\widehat{D}_{0}^3}\left(\widehat{P}_{11}\widehat{P}_{33}-%
\widehat{P}_{13}^2\right) & 0 \\ 
-\frac{DRa}{\widehat{D}_{0}^3}\widehat{P}_{33} & 0 & \frac{DRa}{\widehat{D}%
_{0}^3}\widehat{P}_{13} \\ 
-\frac{D^2}{\widehat{D}_{0}^2}\widehat{P}_{13} & 0 & \frac{D^2}{\widehat{D}%
_{0}^2}\widehat{P}_{11} \\ 
&  & 
\end{pmatrix}
\end{align*}

Consequently, the expression for the velocity $v_{0}^{k}$ takes the form

\begin{equation}
v_{0}^{k}=\frac{1}{\Psi }\left[ \left( 1+\frac{Ra\widehat{P}_{33}}{\widehat{D%
}_{0}^{2}}\right) \delta _{kj}+D\frac{\widehat{P}_{kp}}{\widehat{D}_{0}}%
\varepsilon _{pjn}l^{n}-Ra\frac{\widehat{P}_{kp}}{\widehat{D}_{0}^{2}}%
l^{p}l^{j}+\xi _{kj}\right] \frac{F_{0}^{j}}{\widehat{D}_{0}}.  \label{68.21}
\end{equation}

In order to use these formulas, we have to specify in explicit form the
external force $F_{0}^{j}$. Let us specify it by

\begin{equation}
\vec{F}_{0}=f_{0}\left( \vec{i}\cos \varphi _{1}+\vec{k}\cos \varphi
_{2}\right) ,  \label{01}
\end{equation}

\noindent where 
\begin{equation}
\varphi _{1}=k_{0}z-\omega _{0}t,\varphi _{2}=k_{0}x-\omega _{0}t,
\label{36}
\end{equation}

\noindent or

\begin{eqnarray}
\varphi _{1} &=&\vec{k}_{1}\cdot\vec{x}-\omega _{0}t,\varphi _{2}=\vec{k}%
_{2}\cdot\vec{x}-\omega _{0}t,  \label{37} \\
\vec{k}_{1} &=&k_{0}(0,0,1);\vec{k}_{2}=k_{0}(1,0,0).  \notag
\end{eqnarray}

One can check that $\nabla\cdot\vec{F}_{0}=0$ and $\vec{F}_{0}\cdot
\nabla\times\vec{F}_{0}=0$

Formulas (\ref{01}) and (\ref{37}) allow us to easily make intermediate
calculations, but in the final formulas we obviously shall take $f_{0},k_{0}$
and $\omega _{0}$ as equal to unity, since the external force is
dimensionless and depends only on the dimensionless arguments of space and
time. The force (\ref{01}) is physically simple and can be realized in
laboratory experiments and in numerical simulations.

The force (\ref{01}) can be written in complex form:

\begin{equation}
\vec{F}_{0}=\vec{A}\exp (i\varphi _{1})+ \vec{A}^{\ast }\exp (-i\varphi
_{1})+\vec{B}\exp (i\varphi _{2})+\vec{B}^{\ast }\exp (-i\varphi _{2}),
\label{39}
\end{equation}

\noindent where $\vec{A}$ and $\vec{B}$ have the forms

\begin{equation}
\vec{A}=\frac{f_{0}}{2}\vec{i},\vec{B}=\frac{f_{0}}{2}\vec{k}.  \label{40}
\end{equation}

The effect of the operator $\widehat{D}_{0}$ on the proper function $\exp [i(%
\vec{k}\cdot \vec{x}-\omega t)]$ has obviously the form

$\widehat{D}_{0}\exp[i(\vec{k}\cdot\vec{x}-\omega t)]=\widehat{D}_{0}(\omega
,\vec{k})\exp[i(\vec{k}\cdot\vec{x}-\omega t)]$, where $\widehat{D}%
_{0}(\omega ,\vec{k})$ is

\begin{equation}
\widehat{D}_{0}(\omega ,\vec{k})=i(\omega +\vec{k}\cdot\vec{W})+k^{2}.
\label{68.23}
\end{equation}

>From this it follows that

\begin{equation}
\widehat{D}_{0}(\omega ,-\vec{k}_{1})=i(\omega -\vec{k} _{1}\cdot\vec{W}%
)+k_{1}^{2},  \label{68.24}
\end{equation}

\begin{equation}
\widehat{D}_{0}^{\ast }(\omega ,-\vec{k}_{1})=\widehat{D} _{0}(-\omega ,\vec{%
k}_{1}),  \label{68.25}
\end{equation}

\begin{equation}
\widehat{D}_{0}(\omega ,-\vec{k}_{2})=i(\omega -\vec{k} _{2}\cdot\vec{W}%
)+k_{2}^{2},  \label{68.26}
\end{equation}

\begin{equation}
\widehat{D}_{0}^{\ast }(\omega ,-\vec{k}_{2})=\widehat{D} _{0}(-\omega ,\vec{%
k}_{2}).  \label{68.27}
\end{equation}

>From formulas (\ref{68.21}) and (\ref{39}), it follows that the field $%
v_{0}^{k}$ is composed of four terms:

\begin{equation}
v_{0}^{k}=v_{01}^{k}+v_{02}^{k}+v_{03}^{k}+v_{04}^{k},
\end{equation}

\noindent where

\begin{equation}
v_{02}^{k}=\left( v_{01}^{k}\right) ^{\ast },v_{04}^{k}=\left(
v_{03}^{k}\right) ^{\ast }.
\end{equation}

Finally, we introduce the notation

\begin{equation}
\widehat{D}_{0}(\omega _{0},\vec{-k}_{1})=1+i(1-W_{1})\equiv D_{1},
\label{68.30}
\end{equation}

\begin{equation*}
\widehat{D}_{0}(-\omega _{0},\vec{k}_{1})=D_{1}^{\ast },
\end{equation*}

\begin{equation}
\widehat{D}_{0}(\omega _{0},-\vec{k}_{2})=1+i\left( 1-W_{2}\right) \equiv
D_{2},  \label{68.31}
\end{equation}

\begin{equation*}
\widehat{D}_{0}(-\omega _{0},\vec{k}_{2})=D_{2}^{\ast },
\end{equation*}

\noindent where $\left(W_{1},W_{2}\right)\equiv\left(W^{x},W^{y}\right)$.
Taking into account these formulas, we can write down the velocities $%
v_{0}^{k}$ in the form

\begin{equation}
v_{01}^{k}=M_{kj(1)}^{-1}\frac{A^{j}}{D_{1}^{\ast }}e^{i\varphi _{1}},
\label{807}
\end{equation}

\begin{equation}
v_{03}^{k}=M_{kj(2)}^{-1}\frac{B^{j}}{D_{2}^{\ast }}e^{i\varphi _{2}},
\label{811}
\end{equation}

\noindent where

\begin{equation*}
M_{kj(1)}^{-1}=\frac{1}{\Psi_{(1)}}\left[\left(1+\frac{Ra\widehat{P}_{33}}{%
D_{1}^{\ast2}}\right)\delta _{kj}+D\frac{\widehat{P}_{kp}}{D_{1}^\ast}%
\varepsilon_{pjn}l^{n}-Ra\frac{\widehat{P}_{kp}}{D_{1}^{\ast2}}l^{p}l^{j} +
\xi_{kj(1)} \right]
\end{equation*}
and 
\begin{equation*}
M_{kj(2)}^{-1}=\frac{1}{\Psi_{(2)}} \left[\left(1+\frac{Ra\widehat{P}_{33}}{%
D_{2}^{\ast2}}\right)\delta _{kj}+D\frac{\widehat{P}_{kp}}{D_{2}^\ast}%
\varepsilon_{pjn}l^{n}-Ra\frac{\widehat{P}_{kp}}{D_{2}^{\ast2}}l^{p}l^{j} +
\xi_{kj(2)} \right]
\end{equation*}

We can now calculate the Reynolds stresses:

\begin{equation}
\overline{v_{0}^{k}v_{0}^{i}}=2Re\left( \overline{v_{01}^{k}v_{01}^{i\ast }}+%
\overline{v_{03}^{k^\ast }v_{03}}\right),
\end{equation}

\noindent which can be decomposed into two components:

\begin{equation}
\overline{v_{0}^{k}v_{0}^{i}}=T_{(1)}^{ki}+T_{(2)}^{ki},
\end{equation}

\noindent where $T_{(1)}^{ki}$ and $T_{(2)}^{ki}$ can be expressed as
follows:

\begin{equation}
T_{(1)}^{ki}=\overline{v_{01}^{k}v_{01}^{i\ast }}+\overline{v_{01}^{k\ast
}v_{01}^{i}},
\end{equation}

\begin{equation}
T_{(2)}^{ki}=\overline{v_{03}^{k}v_{03}^{i\ast }}+\overline{v_{03}^{k\ast
}v_{03}^{i}}.
\end{equation}

Taking into account Formulas (\ref{807}) and (\ref{811}), we obtain

\begin{equation*}
\widehat{P}_{11}=\frac{1}{2},  \widehat{P}_{33}=\frac{1}{2},  \widehat{P}_{13}=-\frac{1}{2},  \widehat{P}_{i2}=\delta _{i2}.
\end{equation*}

We can write down the components $T_{(1)}^{3i}$ and $T_{(2)}^{3i}$, which
are the ones of interest:

\begin{equation}
T_{(1)}^{31}=\frac{D^{2}\left( D_{1}^{2}+D_{1}^{\ast 2}+Ra\right) }{8|\Psi
|^{2}|D_{1}|^{6}},
\end{equation}

\begin{equation}
T_{(2)}^{31}=\frac{Ra\left( D_{2}^{2}+D_{2}^{\ast 2}+D^{2}\right) }{8|\Psi
|^{2}|D_{2}|^{6}},
\end{equation}

\begin{equation}
T_{(1)}^{32}=-\frac{D^{3}\left( Ra+2|D_{1}|^{2}\right) }{8|\Psi
|^{2}|D_{1}|^{8}},
\end{equation}

\begin{equation}
T_{(2)}^{32}=-\frac{DRa\left( D_{2}^{3}+D_{2}^{\ast 3}+D^{2}\right) }{8|\Psi
|^{2}|D_{2}|^{8}}.
\end{equation}

Finally, using the following relations (we have similar formulas for $D_{2}$
after replacing $W_{1}$ with $W_{2}$):

\begin{equation}
D_{1}=1+i\left( 1-W_{1}\right) ,
\end{equation}

\begin{equation}
\left\vert D_{1}\right\vert ^{2}=1+\left( 1-W_{1}\right) ^{2},
\end{equation}

\begin{equation}
D_{1}^{2}=1-\left( 1-W_{1}\right) ^{2}+2i\left( 1-W_{1}\right) ,
\end{equation}

\begin{equation}
\left\vert D_{1}\right\vert ^{4}=\left[ 1+(1-W_{1})^{2}\right] ^{2},
\end{equation}

\begin{equation}
D_{1}^{3}=1-3\left( 1-W_{1}\right) ^{2}+i\left[ 1-W_{1}-3\left(
1-W_{1}\right) ^{3}\right] ,
\end{equation}

\begin{equation}
\left\vert D_{1}\right\vert ^{6}=\left[ 1+(1-W_{1})^{2}\right] ^{3},
\end{equation}

\begin{equation}
\left\vert D_{1}\right\vert ^{8}=\left[ 1+(1-W_{1})^{2}\right] ^{4},
\end{equation}

\begin{equation}
\Psi _{(1)}=1+\frac{D^{2}}{2D_{1}^{\ast 2}}+\frac{Ra}{2D_{1}^{\ast 2}},
\end{equation}

\begin{equation}
|\Psi _{(1)}|^{2}=\frac{4|D_{1}|^{4}+\left( 2D^{2}+2Ra\right) \left(
D_{1}^{2}+D_{1}^{\ast 2}\right) +2RaD^{2}+Ra^{2}+D^{4}}{4|D_{1}|^{4}}.
\end{equation}

\bigskip

We can then express $T_{(1)}^{31}$, $T_{(2)}^{31}$, $T_{(1)}^{32}$ and $%
T_{(2)}^{32}$:

\begin{equation*}
T_{(1)}^{31}=\frac{D^{2}[2+Ra-2(1-W_{1})^{2}]}{\Xi _{(1)}},
\end{equation*}

\begin{equation*}
T_{(2)}^{31}=\frac{Ra[2+D^{2}-2(1-W_{2})^{2}]}{\Xi _{(2)}},
\end{equation*}

\begin{equation*}
T_{(1)}^{32}=\frac{-D^{3}[Ra+2[1-(1-W_{1})^{2}]]}{\Xi _{(1)}},
\end{equation*}

\begin{equation*}
T_{(2)}^{32}=\frac{-DRa[2+D^{2}-6(1-W_{2})^{2}]}{\Xi _{(2)}},
\end{equation*}

\noindent where

\begin{equation*}
\Xi
_{(1),(2)}=2[D^{4}+Ra^{2}+2D^{2}Ra+4[1+(1-W_{1,2})^{2}]^{2}+(2D^{2}+2Ra)[2-2(1-W_{1,2})^{2}]][1+(1-W_{1,2})^{2}].
\end{equation*}

\section{Large scale instability}

Let us write down in the explicit form the equations for nonlinear
instability:

\begin{equation}
\partial _{T}W_{1}+\nabla _{Z}T_{(1)}^{31}+\nabla _{Z}T_{(2)}^{31}=\Delta
W_{1},  \label{966}
\end{equation}

\begin{equation}
\partial _{T}W_{2}+\nabla _{Z}T_{(1)}^{32}+\nabla _{Z}T_{(2)}^{32}=\Delta
W_{2},  \label{970}
\end{equation}

\noindent where the components $T_{(1)}^{31}$, $T_{(2)}^{31}$, $T_{(1)}^{32}$
and $T_{(2)}^{32}$ of the Reynolds stress tensor are as defined in the
previous section.\newline

One can see that for small values of the variables $W_{1}$ and $W_{2}$,
Equations (\ref{966}) and (\ref{970}) are reduced to linear equations and
describe the linear stage of instability:

\begin{equation}
\partial _{T}W_{1}+a\nabla _{Z}W_{1}+b\nabla _{Z}W_{2}=\Delta W_{1},
\label{982}
\end{equation}

\begin{equation}
\partial _{T}W_{2}+c\nabla _{Z}W_{1}+d\nabla _{Z}W_{2}=\Delta W_{2},
\label{986}
\end{equation}

\noindent where the coeficients $a$, $b$, $c$ and $d$ can be written as

\begin{align}
&a=\frac{a^{\prime}}{\Lambda} & &b=\frac{b^{\prime}}{\Lambda} & &c=\frac{%
c^{\prime}}{\Lambda} & &d=\frac{d^{\prime}}{\Lambda}
\end{align}

\noindent with

\begin{equation}
a^{\prime
}=4D^{6}+64D^{2}+D^{2}Ra^{3}+(2D^{4}-4D^{2})Ra^{2}+(D^{6}+48D^{2})Ra,
\end{equation}

\begin{equation}
b^{\prime }=(D^{2}+4)Ra^{3}+2D^{4}Ra^{2}+(D^{6}-4D^{4}+48D^{2}+64)Ra,
\end{equation}

\begin{equation}
c^{\prime
}=-2D^{7}+16D^{5}-96D^{3}-D^{3}Ra^{3}+(2D^{3}-2D^{5})Ra^{2}-(D^{7}+16D^{3})Ra,
\end{equation}

\begin{equation}
d^{\prime
}=-(D^{3}+2D)Ra^{3}-(2D^{5}+16D)Ra^{2}+(-D^{7}+2D^{5}-48D^{3}+32D)Ra,
\end{equation}

\noindent and

\begin{equation}
\Lambda =4(D^{4}+Ra^{2}+2RaD^{2}+16)^{2},
\end{equation}

\noindent which are the explicit forms of the quite bulky coefficients.
However, these coefficients can be expressed using the internal helicity $%
H_{0}$ of the velocity field $v_{0}$, calculated in Appendix B. $(H_{0}^{2}=%
\frac{[-16D-DRa^{2}+(D^{3}+4D)Ra]^{2}}{4\Lambda })$. Therefore, we can write
the constant coefficients $a,b,c$ and $d$ with respect to $H_{0}$:

\begin{equation*}
a=H_{0}^{2}\frac{4a^{\prime }}{\Pi },b=H_{0}^{2}\frac{4b^{\prime }}{\Pi }%
,c=H_{0}^{2}\frac{4c^{\prime }}{\Pi },d=H_{0}^{2}\frac{4d^{\prime }}{\Pi },
\end{equation*}

\noindent where $\Pi=[-16D-DRa^{2}+(D^{3}+4D)Ra]^{2}$.

Equations (\ref{982}) and (\ref{986}) can then be rewritten:

\begin{equation}
\partial _{T}W_{1}+H_{0}^{2}\frac{4a^{\prime }}{\Pi }\nabla
_{Z}W_{1}+H_{0}^{2}\frac{4b^{\prime }}{\Pi }\nabla _{Z}W_{2}=\Delta W_{1},
\label{904}
\end{equation}

\begin{equation}
\partial _{T}W_{2}+H_{0}^{2}\frac{4c^{\prime }}{\Pi }\nabla
_{Z}W_{1}+H_{0}^{2}\frac{4d^{\prime }}{\Pi }\nabla _{Z}W_{2}=\Delta W_{2}.
\end{equation}

These formulas show that despite the zero helicity of the driving force,
inside the system, an internal helicity is generated as a result of the
joint impact of the Coriolis and buoyancy forces.\ This helicity plays an
important role in the dynamics of the perturbations.

\subsection{Unstable and oscillatory modes in the case of negligible
viscosity $(k<<1)$}

\bigskip

\bigskip In order to find instabilities, we choose the velocity $W_{1},W_{2}$
in the form:

\begin{equation}
W_{1}=\Upsilon _{1}\exp \left[ i\left( kZ-\omega T\right) \right] ,
\end{equation}

\begin{equation*}
W_{2}=\Upsilon _{2}\exp \left[ i\left( kZ-\omega T\right) \right] .
\end{equation*}

Injecting these solutions into (\ref{982}), we obtain the simple system of
equations:

\begin{equation}
\left( ak-\omega \right) \Upsilon _{1}+bk\Upsilon _{2}=0,
\end{equation}

\begin{equation*}
ck\Upsilon _{1}+\left( dk-\omega \right) \Upsilon _{2}=0.
\end{equation*}

Evidently we get a quadratic equation for $\omega $:

\begin{equation*}
\omega^{2}-\left(a+d\right)k\omega +(ad-bc)k^{2}=0,  \label{1074}
\end{equation*}

\noindent which allows us to obtain the dispersion equations for the
different modes.

\subsubsection{Dispersion equation for the unstable mode}

This equation is obtained by searching for solutions of (\ref{1074}) for
which the discriminant is negative, namely, $\left(a-d\right)^{2}+4bc<0$. We
show below two figures representing the area (in gray) of the plane $(D,Ra)$
for which the discriminant is negative, this means that an instability can
appear. The first figure shows the conditions for a negative temperature
gradient and the second figure, for a positive one.

\begin{figure}[h]
\centerline{\includegraphics[width=0.35\textwidth]{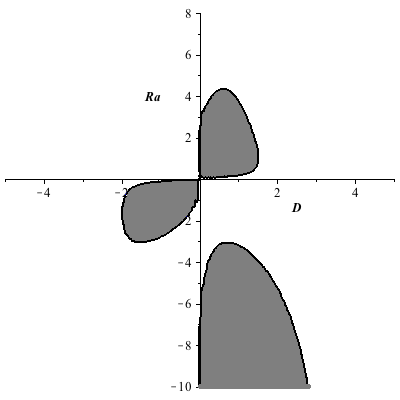}}
\caption{Instability condition with negative temperature gradient}
\end{figure}

\begin{figure}[h]
\centerline{\includegraphics[width=0.35\textwidth]{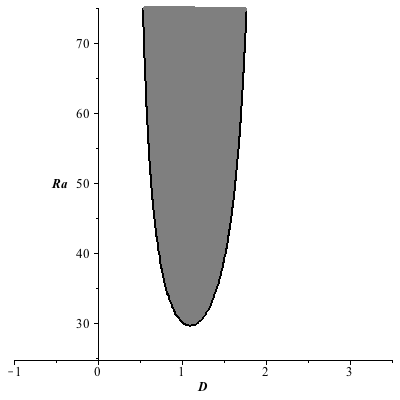}}
\caption{Instability condition with positive temperature gradient}
\end{figure}

Finally, we get

\begin{equation*}
\omega=\omega_{0}\left(D,Ra\right)+i\gamma\left(D,Ra\right)=
\end{equation*}

\begin{equation*}
=4H_{0}^{2}\frac{a^{\prime }+d^{\prime }}{2\Pi }k+i4H_{0}^{2}\frac{\sqrt{-%
\left[ \left( a^{\prime }-d^{\prime }\right) ^{2}+4b^{\prime }c^{\prime }%
\right] }}{2\Pi }k,
\end{equation*}

\noindent where

\begin{equation}
\gamma \left( D,Ra\right) =4H_{0}^{2}\frac{\sqrt{-\left[ \left( a^{\prime
}-d^{\prime }\right) ^{2}+4b^{\prime }c^{\prime }\right] }}{2\Pi }k
\end{equation}

\noindent is the growth rate of the instability. We note that it is
proportionnal to the square of the helicity.

\subsubsection{Dispersion equation for the oscillatory modes}

This equation is obtained by searching for solutions of (\ref{1074}) for
which the discriminant is positive, namely, $\left(a-d\right)^{2}+4bc>0$.

We obtain in this case two oscillatory modes, $\omega_{1}$ and $\omega_{2}$,
which are, respectively, a slow and a fast mode:

\begin{equation}
\omega _{1}=4H_{0}^{2}\frac{a^{\prime }+d^{\prime }-\sqrt{\left[ \left(
a^{\prime }-d^{\prime }\right) ^{2}+4b^{\prime }c^{\prime }\right] }}{2\Pi }%
k,
\end{equation}

\begin{equation}
\omega _{2}=4H_{0}^{2}\frac{a^{\prime }+d^{\prime }+\sqrt{\left[ \left(
a^{\prime }-d^{\prime }\right) ^{2}+4b^{\prime }c^{\prime }\right] }}{2\Pi }%
k.
\end{equation}

It appears that both slow and fast oscillatory frequencies are proportional
to the square of the helicity as well.

\subsection{Unstable and oscillatory modes with viscosity}

In the same way as before, we get the system

\begin{equation}
\left[ k^{2}+i\left( ak-\omega \right) \right] \Upsilon _{1}+ibk\Upsilon
_{2}=0,
\end{equation}

\begin{equation*}
ick\Upsilon _{1}+\left[ k^{2}+i\left( dk-\omega \right) \right] \Upsilon
_{2}=0.
\end{equation*}

We can then get a new quadratic equation for $\omega$:

\begin{equation*}
\omega ^{2}+\left[ 2ik^{2}-\left( a+d\right) k\right] \omega -k^{4}-i\left(
a+d\right) k^{3}+\left( ad-bc\right) k^{2}=0.
\end{equation*}

\subsubsection{Dispersion equation for the unstable mode}

The discriminant of this equation is the same as in the nonviscous case, so
the dispersion equation for the unstable mode has the same condition, namely 
$\left(a-d\right)^{2}+4bc<0$, which leads to:

\begin{equation*}
\omega=\omega_{0}\left(D,Ra\right)+i\gamma\left(D,Ra\right)=
\end{equation*}

\begin{equation*}
=4H_{0}^{2}\frac{a^{\prime }+d^{\prime }}{2\Pi }k+i\left( 4H_{0}^{2}\frac{%
\sqrt{-\left[ \left( a^{\prime }-d^{\prime }\right) ^{2}+4b^{\prime
}c^{\prime }\right] }}{2\Pi }k-k^{2}\right) ,
\end{equation*}

\noindent where

\begin{equation}
\gamma \left( D,Ra\right) =4H_{0}^{2}\frac{\sqrt{-\left[ \left( a^{\prime
}-d^{\prime }\right) ^{2}+4b^{\prime }c^{\prime }\right] }}{2\Pi }k-k^{2}.
\end{equation}

It is to be noted that the growth rate $\gamma \left( D,Ra\right) $ is
maximal for $k_{max}=\frac{H_{0}^{2}}{\Pi }\sqrt{-\left[ (a^{\prime
}-d^{\prime })^{2}+4b^{\prime }c^{\prime }\right] }$, which can be
considered as the characteristic scale of the generated vortex structures.
Below is a figure showing the evolution of $\gamma $ with respect to the
wave number $k$ for $D=Ra=1$.

\begin{figure}[h]
\centerline{\includegraphics[width=0.35\textwidth]{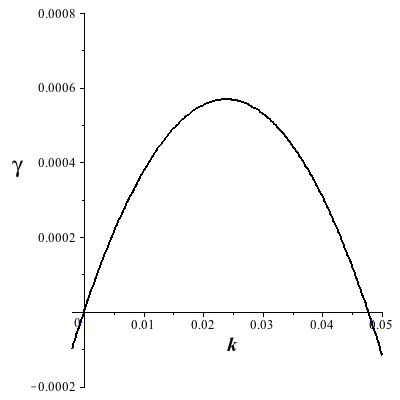}}
\caption{Evolution of the growth rate with respect to $k$}
\end{figure}

It can be noted that if the discrimant is positive, we get an oscillation
with an exponentially decreasing amplitude.

With increasing amplitude, the instability becomes nonlinear and stabilizes.
As a result, nonlinear vortex structures appear. The nonlinear stage of this
instability and the results of numerical simulations will be presented in a
future paper.

\section{Conclusions and discussion of the results}

In this paper, we showed that a large scale instability can appear in a
rotating stratified fluid which is under the impact of a simple small scale
external force (turbulence). The scale of this instability is much larger
than the scale of the external force or turbulence. It is important to
emphasize that, unlike previous papers about large scale instabilities, in
the present paper, there are no special constraints imposed on the external
force. It has a zero helicity and its parity need not be violated; this
means that this is a general force. Nevertheless, the small scale turbulence
under the impact of the Coriolis force and the buoyancy force becomes
helical. This helicity $H_{0}$, finally, is responsible for the generation
of large scale instabilities because the growth rate $\gamma $ is
proportional to $H_{0}^{2} $. The instability itself is oscillating while
its frequency $\omega $ and $\gamma $ have, in principle, the same order.
This means that the instability in the general case is aperiodic. The
frequency of both the stable and unstable oscillations is also proportional
to $H_{0}^{2}$. So we can say that the oscillation modes are inertial
oscillations of the rotating fluid strongly modified by the helicity. There
are two oscillating modes: one slow $\omega _{1}$ and one fast $\omega _{2}$%
. These oscillations decay when the viscosity is taken into account and in
the case of instability, the maximal growth rate is reached at a
characteristic scale of $k_{max}$. Thereby this scale is typical for vortex
structures like Beltrami's runaways. In this paper, the theory of a large
scale instability was constructed using the method of multi-scale
developments, which was proposed in the work of Frisch, She and Sulem \cite
{[15]}. The nonlinear secular equations for the large scale instability were
obtained in the third order of development on a small Reynolds number. In
this paper, we studied in detail the linear stage of the instability and the
conditions of its appearance. It is interesting to note that instability is
possible in the case of both stable and unstable stratifications. Moreover,
that neither the Rayleigh number nor the Taylor number are assumed to be
either big or small: this means that these numbers are out of scheme
parameters. That is the reason why we should state that $Ra<Ra_{cr} $, where 
$Ra_{cr}$ is the critical Rayleigh number for the generation of convective
instability. The unstable stratification is typical for atmosphere dynamics
while the stable one is typical for ocean dynamics. We believe that the
instability which was found in this paper could be applied to the issue of
the generation of large scale vortices in the atmosphere and the ocean, and
to some astrophysical problems as well.

\bigskip

\section*{Appendix}

\appendix

\section{Calculation of the Reynolds stress tensor}

In order to calculate the Reynolds sress, we begin with the general
expression

\begin{equation}
\overline{v_{0}^{k}v_{0}^{i}}=T_{(1)}^{ki}+T_{(2)}^{ki}
\end{equation}

\noindent with 
\begin{equation}
T_{(1)}^{ki}=\overline{v_{01}^{k}v_{01}^{i\ast }}+\overline{v_{01}^{k\ast
}v_{01}^{i}}
\end{equation}
and 
\begin{equation}
T_{(2)}^{ki}=\overline{v_{03}^{k}v_{03}^{i\ast }}+\overline{v_{03}^{k\ast
}v_{03}^{i}.}
\end{equation}

Hence,

\begin{equation}
T_{(1)}^{ki}= M_{kj(1)}^{-1}\frac{A^{j}}{D_{1}^\ast}e^{i\varphi _{1}}\times
M_{it(1)}^{\ast-1}\frac{A^{\ast t}}{D_{1}}e^{-i\varphi _{1}}+
\end{equation}

\begin{equation*}
+M_{kj(1)}^{\ast -1}\frac{A^{\ast j}}{D_{1}}e^{-i\varphi _{1}}\times
M_{it(1)}^{-1}\frac{A^{t}}{D_{1}^{\ast }}e^{i\varphi _{1}},
\end{equation*}

and $T_{(2)}^{ki}$ has a similar expression.\newline

Taking into account that only the components $A_{1},A_{1}^{\ast },B_{3}$ and 
$B_{3}^{\ast }$ of the external force are nonzero, and after some
factorizations, we can write the two contribution of the Reynolds stress
tensor in the following form:

\begin{equation*}
T_{(1)}^{ki}=\frac{\left(1+\frac{Ra\widehat{P}_{33}}{D_{1}^{\ast2}}\right)}{%
\vert\Psi_{(1)}\vert^{2}\vert D_{1}\vert^{2}}A_{i}\xi_{kj(1)}^{\ast}A_{j}^{%
\ast}+ \frac{\left(1+\frac{Ra\widehat{P}_{33}}{D_{1}^{2}}\right)}{%
\vert\Psi_{(1)}\vert^{2}\vert D_{1}\vert^{2}}A_{i}^{\ast}\xi_{kj(1)}A_{j}-%
\frac{D}{\vert\Psi_{(1)}\vert^{2}\vert D_{1}\vert^{2}D_{1}^{\ast}}\widehat{P}%
_{i2}A_{1}\xi_{kj(1)}^{\ast}A_{j}^{\ast}
\end{equation*}

\begin{equation*}
-\frac{D}{|\Psi _{(1)}|^{2}|D_{1}|^{2}D_{1}}\widehat{P}_{i2}A_{1}^{\ast }\xi
_{kj(1)}A_{j}+\frac{1}{|\Psi _{(1)}|^{2}|D_{1}|^{2}}(\xi _{kj(1)}A_{j}\xi
_{it(1)}^{\ast }A_{t}^{\ast }+\xi _{kj(1)}^{\ast }A_{j}^{\ast }\xi
_{it(1)}A_{t}).
\end{equation*}

The same calculation for the contribution $T_{(2)}^{ki}$ gives us

\begin{equation*}
T_{(2)}^{ki}=-\frac{\left(1+\frac{Ra\widehat{P}_{33}}{D_{2}^{\ast2}}%
\right)RaB_{3}^{\ast}}{\vert\Psi_{(2)}\vert^{2}\vert D_{2}\vert^{2} D_{2}^{2}%
}B_{k}\widehat{P}_{i3}-\frac{\left(1+\frac{Ra\widehat{P}_{33}}{D_{2}^{2}}%
\right)RaB_{3}}{\vert\Psi_{(2)}\vert^{2}\vert D_{2}\vert^{2}D_{2}^{\ast2}}%
B_{k}^{\ast}\widehat{P}_{i3}+\frac{\left(1+\frac{Ra\widehat{P}_{33}}{%
D_{2}^{\ast2}}\right)}{\vert\Psi_{(2)}\vert^{2}\vert D_{2}\vert^{2}}%
B_{k}\xi_{it(2)}^{\ast}B_{t}^{\ast}+
\end{equation*}

\begin{equation*}
+\frac{\left(1+\frac{Ra\widehat{P}_{33}}{D_{2}^{2}}\right)}{%
\vert\Psi_{(2)}\vert^{2}\vert D_{2}\vert^{2}}B_{k}^{\ast}\xi_{it(2)}B_{t}+%
\frac{2Ra^{2}B_{3}B_{3}^{\ast}}{\vert\Psi_{(2)}\vert^{2}\vert D_{2}\vert^{6}}%
(\widehat{P}_{i3}\widehat{P}_{k3})-\frac{RaB_{3}}{\vert\Psi_{(2)}\vert^{2}%
\vert D_{2}\vert^{2}D_{2}^{\ast2}}(\widehat{P}_{k3}\xi_{it(2)}^{\ast}B_{t}^{%
\ast}+\widehat{P}_{i3}\xi_{kj(2)}^{\ast}B_{j}^{\ast})-
\end{equation*}

\begin{equation*}
-\frac{RaB_{3}^{\ast }}{|\Psi _{(2)}|^{2}|D_{2}|^{2}D_{2}^{2}}(\widehat{P}%
_{k3}\xi _{it(2)}B_{t}+\widehat{P}_{i3}\xi _{kj(2)}B_{j})+\frac{1}{|\Psi
_{(2)}|^{2}|D_{2}|^{2}}(\xi _{kj(2)}B_{j}\xi _{it(2)}^{\ast }B_{t}^{\ast
}+\xi _{kj(2)}^{\ast }B_{j}^{\ast }\xi _{it(2)}B_{t}).
\end{equation*}

\section{Calculation of the helicity}

The driving force has no helicity, but the joint action of the external
force, Coriolis force, and the buoyancy give the internal helicity.

The general \ helicity of the velocity field $v_{0}$ is expressed by

\begin{equation}
H=\overline{\vec{v_{0}}\cdot\nabla\times\vec{v_{0}}}=\overline{\vec{v}%
_{01}\cdot\nabla\times\vec{v}_{01}^{\ast}}+\overline{\vec{v}%
_{01}^{\ast}\cdot\nabla\times\vec{v}_{01}}+
\end{equation}

\begin{equation*}
+\overline{\vec{v}_{03}\cdot \nabla \times \vec{v}_{03}^{\ast }}+\overline{%
\vec{v}_{03}^{\ast }\cdot \nabla \times \vec{v}_{03}}=H_{(1)}+H_{(2)},
\end{equation*}

\noindent where we choose $H_{(1)}$ and $H_{(2)}$ such that

\begin{equation*}
H_{(1)}=\overline{\vec{v}_{01}\cdot\nabla\times\vec{v}_{01}^{\ast}}+%
\overline{\vec{v}_{01}^{\ast}\cdot\nabla\times\vec{v}_{01}}
\end{equation*}

\noindent and

\begin{equation*}
H_{(2)}=\overline{\vec{v}_{03}\cdot\nabla\times\vec{v}_{03}^{\ast}}+%
\overline{\vec{v}_{03}^{\ast}\cdot\nabla\times\vec{v}_{03}},
\end{equation*}

\noindent and in indicial notation:

\begin{equation*}
\vec{v_{0}}\cdot \nabla \times \vec{v_{0}}=v_{0}^{k}\varepsilon
_{kui}\partial _{u}v_{0}^{i}.
\end{equation*}

We must calculate $H$ with

\begin{equation}
H_{(1)}= M_{kj(1)}^{-1}\frac{A^{j}}{D_{1}^\ast}e^{i\varphi
_{1}}\times\varepsilon_{kui}\partial_{u} \left( M_{it(1)}^{\ast-1}\frac{%
A^{\ast t}}{D_{1}}e^{-i\varphi _{1}}\right)+
\end{equation}

\begin{equation*}
+M_{kj(1)}^{\ast-1}\frac{A^{\ast j}}{D_{1}}e^{-i\varphi _{1}}\times
\varepsilon_{kui}\partial_{u} \left( M_{it(1)}^{-1}\frac{A^{t}}{D_{1}^\ast}%
e^{i\varphi _{1}}\right)
\end{equation*}

$H_{(2)}$ is calculated in the same way, by replacing $v_{01}$ with $v_{03}$.

We finally obtain

\begin{equation*}
H=\frac{%
D(1-W_{1})[[1+(1-W_{1})^{2}][2Ra-4[1+(1-W_{1})^{2}]]-2Ra[1-3(1-W_{1})^{2}]-Ra^{2}]%
}{%
[D^{4}+Ra^{2}+2D^{2}Ra+4[1+(1-W_{1})^{2}]^{2}+(2D^{2}+2Ra)[2-2(1-W_{1})^{2}]][1+(1-W_{1})^{2}]^{2}%
}+
\end{equation*}

\begin{equation*}
+\frac{DRa(1-W_{2})\left[ 2-6(1-W_{2})^{2}+D^{2}\right] }{%
[D^{4}+Ra^{2}+2D^{2}Ra+4[1+(1-W_{2})^{2}]^{2}+(2D^{2}+2Ra)[2-2(1-W_{2})^{2}]][1+(1-W_{2})^{2}]^{2}%
}.
\end{equation*}

After linearization:

\begin{equation*}
H=\frac{%
16D^{5}+128D^{3}-256D-DRa^{4}-2D^{3}Ra^{3}-(D^{5}+8D^{3}+96D)Ra^{2}-(8D^{5}+32D^{3}-256D)Ra%
}{\Lambda}W_{1}
\end{equation*}

\begin{equation*}
+\frac{{(D^{3}+8D)Ra^{3}+(2D^{5}+8D^{3}+32D)Ra^{2}+(D^{7}+80D^{3})Ra}}{%
\Lambda}W_{2}
\end{equation*}

\begin{equation*}
+\frac{-16D-DRa^{2}+(D^{3}+4D)Ra}{2\sqrt{\Lambda }}.
\end{equation*}

Where we recall that $\Lambda=4(D^{4}+Ra^{2}+2D^{2}Ra+16)^{2}$.\newline

One can note that for small perturbations $(W_{1},W_{2})$, the helicity
approaches the constant:

\begin{equation*}
H_{0}=\frac{-16D-DRa^{2}+(D^{3}+4D)Ra}{2\sqrt{\Lambda }},
\end{equation*}

\noindent which can be considered as the internal helicity of the field $%
v_{0}$ when there are no perturbations.

\end{document}